# Design rules for dislocation filters

T. Ward,[1)] A.M Sánchez,[1)] M. Tang,[2)] J. Wu,[2)] H. Liu,[2)] D.J. Dunstan,[3)] and R. Beanland[1,a)]

[1]*Department of Physics, University of Warwick, Coventry, CV4 7AL, UK*

[2]*Department of Electronic & Electrical Engineering, University College London, Torrington Place, London WC1E 7JE, UK*

[3]*School of Physics and Astronomy, Queen Mary, University of London, London E1 4NS. UK*

The efficacy of strained layer threading dislocation filter structures in single crystal epitaxial layers is evaluated using numerical modeling for (001) face-centred cubic materials, such as GaAs or $Si_{1-x}Ge_x$, and (0001) hexagonal materials such as GaN. We find that threading dislocation densities decay exponentially as a function of the strain relieved, irrespective of the fraction of threading dislocations that are mobile. Reactions between threading dislocations tend to produce a population that is a balanced mixture of mobile and sessile in (001) cubic materials. In contrast, mobile threading dislocations tend to be lost very rapidly in (0001) GaN, often with little or no reduction in the immobile dislocation density. The capture radius for threading dislocation interactions is estimated to be approx. 40nm using cross section transmission electron microscopy of dislocation filtering structures in GaAs monolithically grown on Si. We find that the minimum threading dislocation density that can be obtained in any given structure is likely to be limited by kinetic effects to approx. $10^4 – 10^5$ cm$^{-2}$.

1. ## INTRODUCTION

Growth of mismatched heteroepitaxial layers on single crystal substrates has been an active field of research for over forty years.[1] The inevitable presence of dislocations that thread through the deposited thin film – and the compromised material properties that result – has been a persistent problem, a significant issue in the incorporation of III-V materials with silicon[2] and III-nitride epitaxial layers on sapphire substrates.[3,4] The technological relevance of this problem ensures continued interest in this field[5] as new growth techniques and materials are investigated. Here, we reconsider the role of threading dislocations (TDs) in the accommodation of misfit strain $\varepsilon$ and use numerical modelling to evaluate strategies that can be employed to reduce their density $\rho_{TD}$ in high misfit systems.

---

[a)] Author to whom correspondence should be addressed. Electronic mail: r.beanland@warwick.ac.uk.

In low-misfit layers, mobile TDs provide the mechanism for strain relief since they leave a misfit dislocation in their wake as they move (Fig. 1). As noted by Matthews and co-workers,[1] the system can be described physically as a balance of forces on a threading dislocation, in which the force driving the TD onwards due to the strain in the layer, $F_m \propto \varepsilon h$, is balanced by the line tension of the misfit dislocation $F_L \propto \ln(h)$. This leads to the concept of a critical thickness $h_c$ above which relaxation is energetically favoured, and above $h_c$ an equilibrium can be defined using the same concept of a force balance. The problem can also be couched in terms of energy;[6] the two approaches are formally equivalent and although there are many errors in the literature definitive reviews have been provided by Fitzgerald[7] and Dunstan,[8] for example. As pointed out by Dunstan[8] the logarithmic term has only a small influence and a good rule-of-thumb for equilibrium strain is given by $\varepsilon h \approx A$, a constant (experimentally, in the $In_xGa_{1-x}As$ system $A \approx 0.2$ nm[8]). One might therefore imagine a system in which, as the layer thickness $h$ increases during crystal growth, TD movement proceeds until the strain drops to an equilibrium value $\varepsilon \approx A/h$.

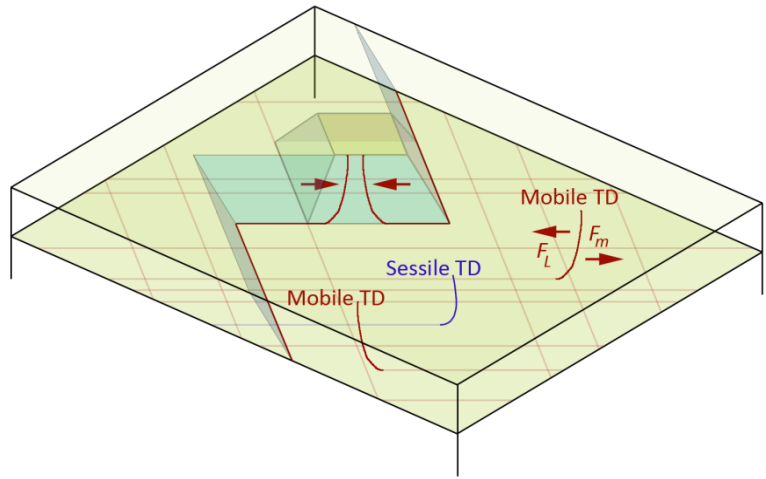

Figure 1. Mobile (glissile) threading dislocations in an epitaxial layer are able to glide on inclined planes, producing an array of misfit dislocations. In cubic (001) layers these are (111) planes, illustrated by the truncated pyramid. Those with opposite Burgers vectors experience an attractive force to each other and annihilate when they meet. Immobile (sessile) TDs do not have an inclined glide plane. They can react with a mobile dislocation to produce a new mobile TD.

In practice, strained layers above $h_c$ do not attain this equilibrium. The strain fields of the misfit dislocations, formed by moving TDs, act upon any TD that moves perpendicular to them and impose additional attractive or repulsive forces that can block[9] or trap[10,11] the mobile TD. Relaxation is thus limited by this effect, rather than the balance of forces on an isolated dislocation; it has been estimated that this increases the strain-thickness product $\varepsilon h$ above equilibrium by approx. 20%.[10] Furthermore when $\rho_{TD}$ is low there is a limit on the amount of strain that can be relieved, $\varepsilon_r$; dislocation velocities are finite[7] and the amount of strain relieved by the movement of pre-existing dislocations can be almost imperceptible. Then it is necessary for dislocation multiplication to take place, and simple arguments show that this requires approx. $\varepsilon h \approx 0.8$ nm.[12] These effects are illustrated in Fig. 2, which shows equilibrium, trapping and multiplication curves as a function of layer thickness. Different relaxation behaviours are shown; (i) dislocation kinetics/multiplication limited relaxation, for highly perfect material with low initial $\rho_{TD}$[13] and (ii) trapping-limited relaxation, expected in a system with a high density of mobile TDs.[10] In both these cases repulsive dislocation interactions begin to dominate at some point (particularly if dislocation pile-ups are produced by multiplication sources[11,14]), leading to a finite residual strain, i.e. a work-hardening effect.[8,15] Similar effects are seen in nanoscale deformation experiments.[16] In the case of high misfit layers, (iii), the majority of relaxation occurs very early in growth. The interfacial energy is high and 3D ('island') growth is usual. Relaxation occurs through different mechanisms to those shown in Fig. 1, and the resulting dislocation densities are very high. We show in Fig. 2 the possibility that the remaining strain may be below the planar layer equilibrium value, although as growth proceeds and the layer becomes continuous any further relaxation must follow the constraints on planar layers.

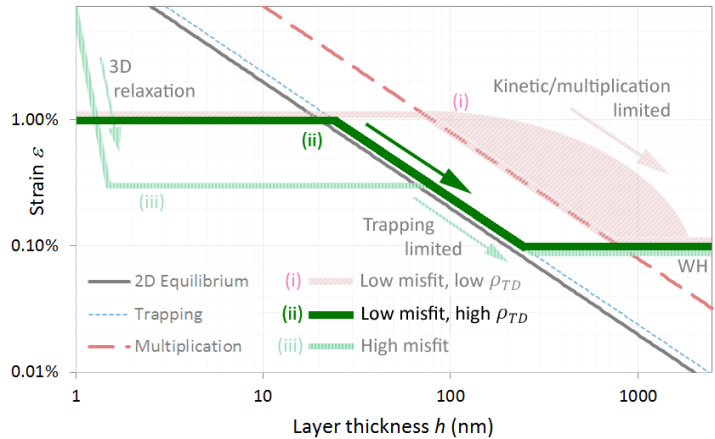

Figure 2. Illustration of the accommodation of misfit strain in systems with differing levels of misfit and threading dislocation density as a function of layer thickness. Straight lines on this log-log plot show the limits on relaxation due to equilibrium (black line), trapping-limited (blue dotted line), and multiplication (red dashed line). For low misfits relaxation can be limited by (i) multiplication and/or kinetics or (ii) TD blocking/trapping. High misfit materials (iii) may relax by different mechanisms and only follow the same rules once they are planar layers. For all materials a residual strain due to work hardening (WH) may be expected at large thicknesses.

This understanding provides several strategies that can be used to obtain a low $\rho_{TD}$. For low misfit materials the most obvious is to complete the relaxation as the first step, and then to grow a succession of layers, each with $\varepsilon h$ < 0.2 nm by appropriate choice of layer composition and thickness. Using positive and negative strains this can be continued indefinitely in a 'strain balanced' structure.[17] A low initial $\rho_{TD}$ can be maintained by preventing relaxation by restricting the efficacy of mobile dislocations. This is the idea behind growth on mesas [18] or on highly perfect substrate material[13,19]. However, neither of these produce a difference in in-plane lattice parameter in the epitaxial layers. A third approach has been to use slowly graded compositions over layers with thicknesses >1μm,[20] in which dislocations are not pinned at interfaces and work-hardening effects are reduced, providing relatively efficient relaxation. Threading dislocation densities $\rho_{TD}$ below <$10^5$ cm$^{-2}$ can be obtained[21] even after a significant change in lattice parameter. However, none of these approaches tackle the problem of materials with completely different structures, or cases where gradations in composition cannot be produced. In these systems a very high $\rho_{TD}$ (up to or exceeding $>10^{10}$ cm$^{-2}$) is common

In what follows we consider the strategies that may be used to reduce TD density at the crystal surface, using a combination of simple numerical modelling and analytic models. The main application of this approach lies in cases where a high threading dislocation density cannot be avoided (e.g. epitaxial growth of very different structures or materials with high lattice mismatch (>4%). Although this topic has been a subject of investigation for some time there has been relatively little attention paid to the influence of crystallography on behaviour and this is emphasised here. Simple growth of thick epitaxial layers is considered first, followed by the use of 'dislocation filter' structures.

## 2. THREADING DISLOCATIONS IN THICK HIGH-MISFIT LAYERS

It has been known for some time[22-24] that although the initial value of $\rho_{TD}$ is very large in high misfit layers, it begins to drop very rapidly once planar growth takes hold, and continues to reduce as film thickness increases[25,26]. Therefore, an obvious strategy to reduce $\rho_{TD}$, without changing lattice parameter, is simply to grow a 'buffer' layer of sufficient thickness to obtain a sufficiently low defect density at the surface. From the very start[24] it became apparent that such thick layers can lead to problems due to differences in thermal expansion coefficient, e.g. wafer bowing or cracking in thick layers under tension.[27] These problems provide an incentive to develop buffer layers that are as thin as possible (see section 3). First however we focus on this 'natural' decrease in $\rho_{TD}$ with thickness, which must be the result of reactions between the threading dislocations.[20] We consider the system using physical arguments before using this understanding to construct a simple numerical model.

An attractive force per unit length ($F_i/L \propto 1/r$) is present between TDs that would reduce energy if they were to react with each other. This interaction between TDs is often described in terms of a capture radius $R$,[28-31] such that dislocations that pass within a distance $R$ of each other and have an energetically favourable reaction are deemed to meet and react; i.e. the attractive force $F_i$ is sufficient to overcome the energy barrier for dislocation glide.[32,33] As pointed out by Fertig,[28,34,35] the region of attraction between dislocations is highly stress-dependent, reaching a maximum value of several hundred times the magnitude of the Burgers vector $b$ (i.e. tens to hundreds of nm) when the layer is close to the critical strain-thickness product. Furthermore, the region is far from circular and[36] depends on the Burgers vectors of the two dislocations. The use of a simple radius $R$ can only be taken as some indicative measure of the likelihood of interaction. To overcome this problem we thus describe the area over which an interaction will occur to be given by area of interaction = $CR^2$, where $C$ is a constant (if the region were circular, $C = \pi$).

It is apparent that if $R$ is constant, all TDs are fixed in their position and lie perpendicular to the surface, after an initial drop in $\rho_{TD}$ (while dislocations that fall within $R$ react and combine), one would expect no change in $\rho_{TD}$ as thickness increases. Experimental measurements show that $\rho_{TD}$ does decrease; therefore either the TDs must move, or $R$ must increase with thickness. Mathis et al.[31] pointed out that dislocations with an inclined line direction have an intersection with the surface that moves laterally during growth, while Zogg[23] noted that they present an obstacle whose size increases in proportion with $h$ for another inclined dislocation on a different glide plane. Another possibility is that the strain state becomes closer to the critical strain-thickness product as thickness increases (Fig. 2), also resulting in an increase in $R$.[36] As shown by Zogg[23] and Wang[33] when $R \propto h$, the threading dislocation density $\rho_{TD}$ decreases in proportion to the square of the thickness. We now develop a similar argument here, using an interaction radius $R = Kh$.

During growth, as the layer thickness increases by $\delta h$, let the interaction radius for each TD increase by an amount $\delta R$, increasing its interaction area by $\delta A = 2CR\delta R$ (if the interaction radius were circular, $2C = 2\pi$). Since there are $\rho_{TD}$ dislocations per unit area, the total increase in interaction area is $2CR\rho_{TD}\delta R$ per unit area. This will produce $2CR\rho_{TD}^2\delta R$ new encounters per unit area; if each encounter has a probability $q$ of annihilation for one of the TDs and using $R = Kh$, the change in threading dislocation density will be

$$\delta\rho_{TD} = 2CK^2 q \rho_{TD}^2 h \delta h \tag{1}$$

and integration yields

$$\rho_{TD} = \frac{\rho_{TD}^{(0)}}{\left(1 + \rho_{TD}^{(0)} CqK^2 h^2\right)}. \tag{2}$$

Where $\rho_{TD}^{(0)}$ is the initial TD density at $h = 0$. At large thicknesses, when the second term in the denominator is large, the threading dislocation density should follow an inverse square law with layer thickness, i.e.°

$$\rho_{TD} \approx \frac{1}{Bh^2}, \tag{3}$$

Where $B = CqK^2$. Interestingly, the dislocation density at large thicknesses is independent of the initial threading dislocation density. In any given case, the parameter $B$ depends on the material and its crystallography: the parameter $C$ depends upon the shape of the interaction area; $q$ depends upon the allowed reactions between dislocations; and $K$ may depend on several factors including elastic constants and strain state. In order to obtain an estimate of the efficacy of threading dislocation density reduction we used a simple numerical model in which $K = 1$ and $C = \pi$ (i.e. we make the usual compromise in order to obtain a simple model that will give an order-of-magnitude estimate). Statistically, the most common reactions between TDs must involve those with different Burgers vectors (which we call here $\mathbf{b}_1$ and $\mathbf{b}_2$). These can combine to produce a new dislocation with Burgers vector $\mathbf{b}_3 = \mathbf{b}_1 + \mathbf{b}_2$ and reduce $\rho_{TD}$ by a 2 → 1 reaction. Ignoring core energies and other factors, the reaction is energetically favourable when

$$\mathbf{b}_3 = \mathbf{b}_1 + \mathbf{b}_2; \qquad b_3^2 < b_1^2 + b_2^2, \qquad (4)$$

giving an attractive force between the dislocations, while a repulsion exists between those with $b_3^2 > b_1^2 + b_2^2$. Dislocations with opposite Burgers vectors ($\mathbf{b}_1 = -\mathbf{b}_2$) have no reaction product, i.e. a 2 → 0 annihilation reaction. However, in some cases energetically favourable 2 → 2 reactions are also possible, changing dislocation types while not reducing $\rho_{TD}$. The reactions that can occur are dictated by the crystal system and here we consider two systems relevant to group IV and III-V semiconductor materials, i.e. (001) face-centred cubic (fcc) and (0001) hexagonal layers (Fig. 3).

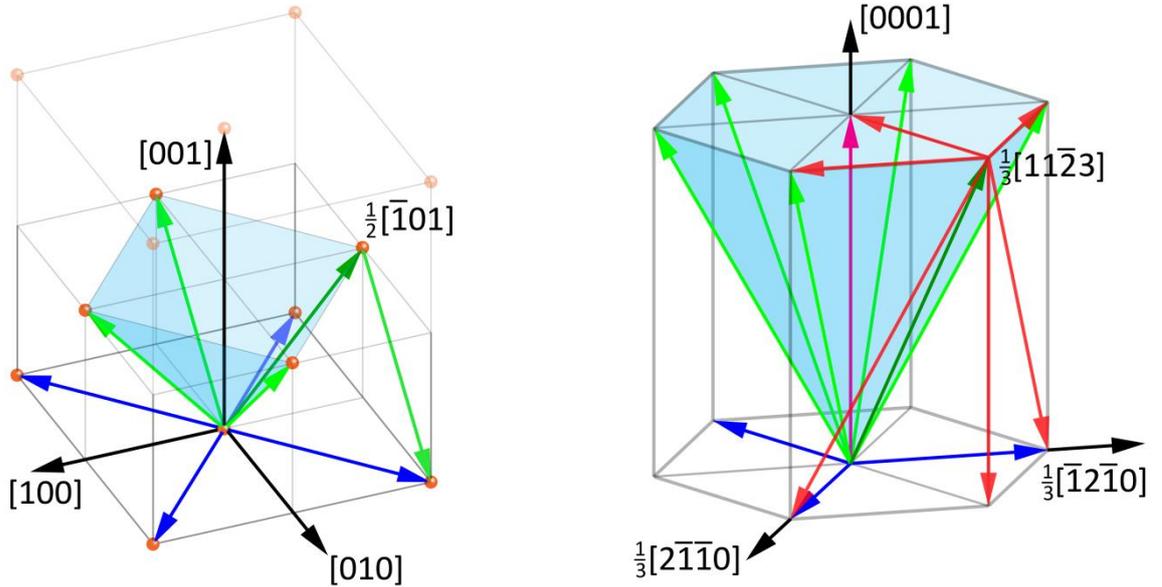

Figure 3. (a) ½<110> Burgers vectors in the fcc system. In (001) layers the green arrows represent Burgers vectors of glissile TDs and the blue arrows represent those of sessile TDs. Four vectors with [00-1] components are not shown. A glissile TD (e.g. ½[-101]) can react with four other TDs (red) to give a ½<110> reaction product. (b) Burgers vectors in the hexagonal system, *a*-type (blue), *c*-type (magenta) and (*a+c*)-type (green). Seven vectors with [000-1] components are not shown. In (0001) layers only the (*a+c*)-type are glissile (e.g. 1/3[11-23]) and this can react with six different TDs (red) to give a single product.

The model was comprised of a square field in which dislocations were distributed at random, initially with Burgers vectors also assigned randomly amongst all possible types (12 types for fcc cubic, 20 types for hexagonal). Lattice parameters corresponding to GaAs ($a = 0.56535$ nm) and GaN ($a = 0.3189$, $c = 0.5186$)[37] were used. As shown in Fig. 3, some dislocation types do not have an inclined glide plane and are thus sessile. This raises the question of whether a pair of sessile TDs should be allowed to react in the model, even if the reaction is energetically favourable since they are, in principle, immobile. Interestingly, as inspection of Fig. 3 shows, in (001) fcc layers this question does not arise, since all triple points must join two glissile dislocations to one sessile dislocation and therefore at least one of the dislocations can move in any conceivable configuration. In contrast, in (0001) hexagonal layers there are reactions between immobile TDs that would reduce energy – if the dislocations could move, to react. In particular, TDs with Burgers vectors in the basal plane can react following the rule $\mathbf{a}_1 + \mathbf{a}_2 = \mathbf{a}_3$, when the angle between $\mathbf{a}_1$ and $\mathbf{a}_2$ is 60°. We have thus modelled two scenarios for GaN: a) where TDs < $R$ with an energetically favourable reaction combine, even if they are both sessile; and b) only allowing reactions to occur when at least one TD is mobile.

We consider first the scenario where all favourable reactions take place, irrespective of the ability of a dislocation to move on a glide plane or not. In each iteration of the model, reactions between the closest dislocations falling in a circular capture area were deemed to occur, according to Eq. (4). For simplicity, neighbouring TDs with repulsive forces did not impede the reacting TDs, even if they were in between the reacting pair. Each run used a population of $10^4$-$10^5$ TDs with length scales chosen to give $\rho_{TD}^{(0)}$ from $10^5$ cm$^{-2}$ to $10^{10}$ cm$^{-2}$. The result is shown in Fig. 4, together with a curve for Eq. (2) for $\rho_{TD}^{(0)} = 10^9$ cm$^{-2}$ and the limiting curve of Eq. (3). The equations are (unsurprisingly) a good fit to the simulated data. The main difference between the two materials systems is that the reactions occur more rapidly in GaN. This reflects the greater probability of reactions that may occur in the hexagonal system. In fcc materials with equal populations of all Burgers vector types the probability of annihilation $q = 1/4$, giving $B = \pi/4$ in our simple model. In contrast for GaN there is a cascade of probabilities; $c$-type dislocations are most likely to react, with $q = 0.35$, followed by ($a+c$) type with $q = 0.2$ and $a$-type with $q = 0.1$. This is also in agreement with experimental observations in which $c$-type dislocations are rarely observed.[38]

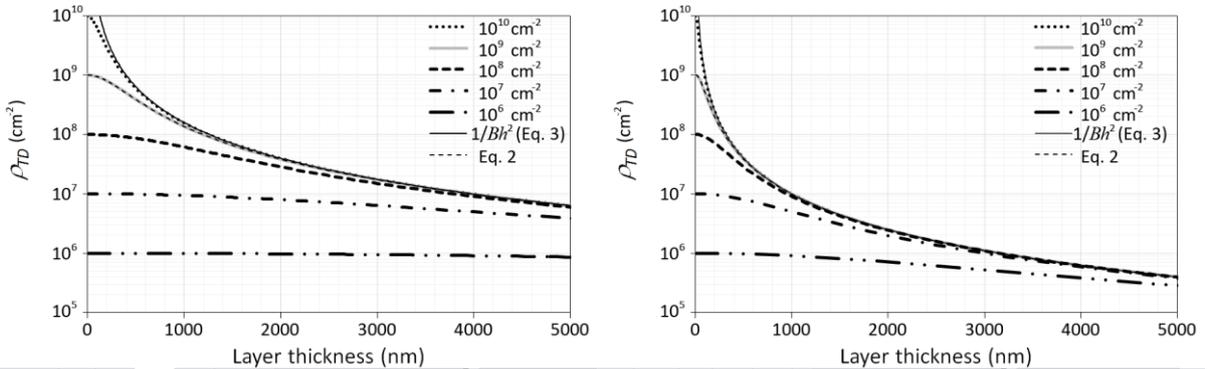

Figure 4. Decrease in $\rho_{TD}$ as a function of film thickness (a) for (001) fcc layers and (b) for (0001) GaN obtained with a simple numerical model in which all energetically favourable reactions are allowed within a circular capture radius $R = h$.

In the case of (0001) GaN layers however, $\rho_{TD}$ is found to be much higher than predicted by Fig. 4b, typically of the order of $10^9 - 10^{10}$ cm$^{-2}$.[31,39] This implies that sessile TDs are unable to react even though a reaction may be energetically favourable. The behaviour that results when this condition is imposed is shown in Fig. 5a. Here, there is an initial decrease, that is well described by Eq. (3) but with a weaker interaction strength in comparison with Fig. 4b, followed by a constant $\rho_{TD}$ that is roughly half the initial value $\rho_{TD}^{(0)}$. As noted by Mathis et al.,[31] the origin of this effect lies in a reaction between pairs of glissile (*a+c*) dislocations. Half of the possible encounters between glissile TDs result in 2 → 2 reactions, e.g. (*a+c*) + (*a-c*) = *a* + *a*. This gives no decrease in $\rho_{TD}$ and is a disaster for threading dislocation density reduction; eventually all mobile TDs are converted into immobile TDs that are effectively unreactive. This becomes apparent when the populations of glissile and sessile dislocations are examined separately (Fig. 6); the glissile TD population drops following a law similar to Eq. (2), but in the early stages of growth the density of sessile TDs actually increases, as pairs of mobile TDs are converted into sessile TDs. Only after the mobile TDs are significantly depleted does the density of sessile TDs begin to drop, although it only manages to reach its original value at the final stage when all mobile TDs are eliminated. This behaviour is consistent with experimental observations, which show that the majority of TDs are sessile *a*-type.[40,41]

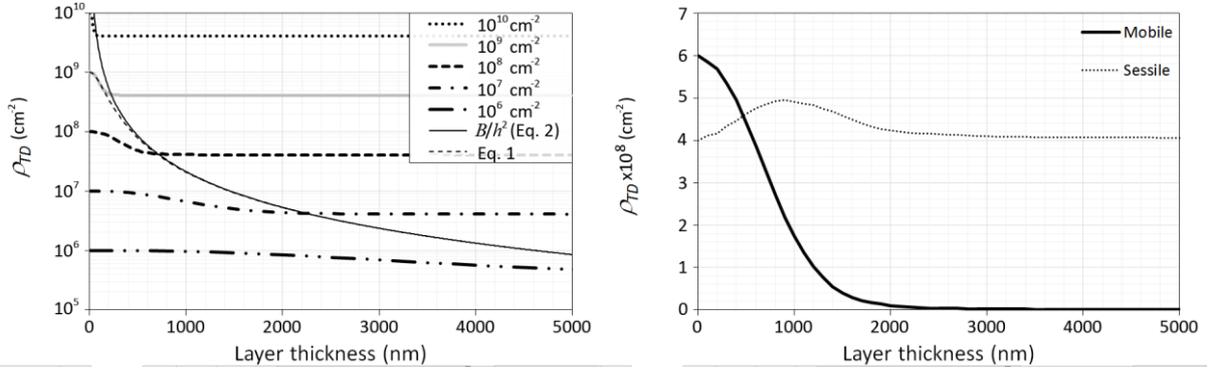

Figure 5. (a) Decrease in $\rho_{TD}$ as a function of film thickness for (0001) GaN layers obtained with a simple numerical model where at least one of the TDs in any reaction must be glissile within a circular capture radius $R = h$. (b) the change in the populations of glissile and sessile TDs for $\rho_{TD}^{(0)} = 10^9$ cm$^{-2}$.

As noted by Romanov,[42] If TDs have line directions that are inclined at an angle $\theta$ from perpendicular this leads to a change in their intersection with the surface as a layer is grown. For an incremental change in layer thickness $\delta h$ the interaction area of each TD moves by a distance $\delta s$ and generates a new interaction area $2R\ \delta s = 2R \tan(\theta)\ \delta h$. The resulting change in TD density is

$$\delta \rho_{TD} = -2Rq \tan(\theta) \rho_{TD}^2 \delta h, \qquad (5)$$

giving

$$\rho_{TD} = \frac{\rho_{TD}^{(0)}}{1 + 2Rq \tan(\theta) \rho_{TD}^{(0)} h}. \qquad (6)$$

Which at large *h* approximates to

$$\rho_{TD} \approx \frac{1}{\beta h}. \tag{7}$$

i.e. a 1/$h$ dependence with thickness.[42] However, this reduction is smaller than that given by Eq. (3) since the coefficient $\beta$ is the same order of magnitude as $R$, i.e. ~$10^{-5}$ cm. Fig. 6 shows the results of numerical modelling in which all TDs move a lateral distance corresponding to an inclination of 45° for fcc cubic layers. TDs exiting the field at one edge entered at the opposite edge at a random position. There is essentially no impact on TD densities of $10^8$ cm$^2$ and below. The reduction in $\rho_{TD}$ for (0001) GaN was similarly poor (not shown here). These simulations show that the use of a simple buffer layer is unlikely to reduce threading dislocation densities to technologically desirable levels (i.e. < $10^5$ cm$^{-2}$) without using extremely thick layers if dislocations do not move laterally to any significant extent. We now consider more complex epitaxial structures that contain 'dislocation filters', with the aim of producing a significantly reduced $\rho_{TD}$ in relatively thin buffer layers.

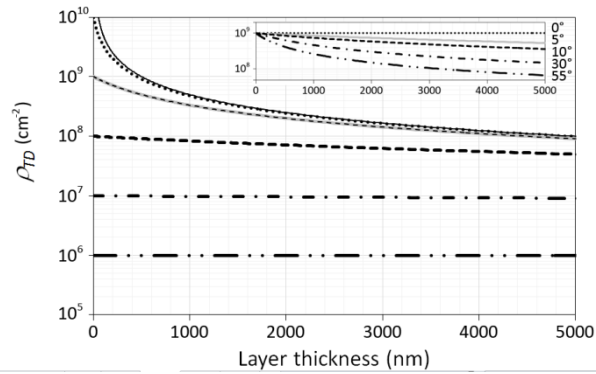

Figure 6. Decrease in $\rho_{TD}$ as a function of film thickness for (001) fcc cubic layers obtained with a simple numerical model with TDs inclined at 45° to the surface. The insert shows the effect of inclination angle from 0° to 55°.

## 3. THREADING DISLOCATION FILTERS

In GaN the use lateral overgrowth[43] to reduce $\rho_{TD}$ has been the subject of much work and will not be considered here – our main concern is whether a single epitaxial growth run is capable of producing low $\rho_{TD}$ through the use of layers with different compositions and thicknesses. The previous section shows that movement is essential to encourage TDs to react and recombine, and it has long been appreciated that using additional misfit strain in 'dislocation filter' layers is one of the principal means to achieve this.[29,42,44,45] A similar realization has informed recent work in nanoscale strengthening mechanisms – so-called 'mechanical annealing'.[46] Since we expect a growing strained layer with a high initial $\rho_{TD}$ to be trapping-limited and thus to always have a strain state close to the critical strain-thickness product (Fig. 2) we interpret this to mean that $R$ is constant and large, typically a few hundred times $b$, i.e. a few tens of nm.

Analytically, similar arguments can be used to those in the previous section to obtain a differential equation that captures the essence of the behaviour for a mixture of mobile and sessile TDs with fixed interaction radius $R$. If the fraction of mobile TDs is $m$ and they move with velocity $v$, over a small time interval $\delta t$ they sweep out an interaction area $2Rv\delta t$, and the change in TD density is

$$\delta \rho_{TD} = -2Rmq\rho_{TD}^2 v\delta t . \tag{8}$$

While the increase in the misfit dislocation length per unit area produces a relaxation

$$\delta \varepsilon_r = Dm\rho_{TD} b_{//} v\delta t \tag{9}$$

where $D$ is a constant ($D = ½$ in the case of an orthogonal array; $D = 2/3$ for a triangular array), and $b_{//}$ is the magnitude of the Burgers vector component that relieves misfit strain (i.e. the edge component lying in the interfacial plane). Combining (4) and (5) gives

$$\delta \rho_{TD} = -\frac{2Rq\rho_{TD}\delta\varepsilon_r}{Db_{//}} . \tag{10}$$

Note the absence in Eq. (10) of the factor $m$, the fraction of mobile dislocations. The reduction in threading dislocation density as a function of strain relieved is thus independent of how many dislocations are mobile, and how many are not. Initially this might seem counterintuitive, but it is simply because the same amount of relaxation $\delta\varepsilon_r$ can be produced by many TDs moving a short distance, or a few TDs moving a long distance. Assuming that $R$, $q$ and $D$ are independent of the state of relaxation, integration of Eq. (10) gives the result that the threading dislocation density decreases exponentially with the strain relaxed $\varepsilon_r$, i.e.[29]

$$\rho_{TD}(\varepsilon_r) = \rho_{TD}^{(0)} \exp\left(\frac{-2Rq\varepsilon_r}{Db_{//}}\right), \tag{11}$$

Note that the fractional reduction, $\rho_{TD}/\rho_{TD}^{(0)}$, does not depend upon the initial dislocation density; this is used later in the measurement of interaction radius $R$ in section 4. The fraction of encounters that lead to the removal of a TD, $q$, depends upon the different populations of Burgers vectors. If there are equal numbers of all Burgers vectors we find;

$$\rho_{TD}(\varepsilon_r) \approx \rho_{TD}^{(0)} \exp\left(\frac{-7R\varepsilon_r}{2b}\right) \text{ (cubic)}, \quad \rho_{TD}(\varepsilon_r) \approx \rho_{TD}^{(0)} \exp\left(\frac{-9R\varepsilon_r}{4b}\right) \text{ (hexagonal)} \tag{12}$$

i.e. TD removal due to movement of mobile TDs is expected to be roughly twice as efficient for (001) cubic layers in comparison with (0001) hexagonal materials.

Numerical modelling of such a system is a simple extension of that described in section 3, with the added element of glissile TD movement in response to the misfit stress. Thus, an incremental change in the position of all mobile TDs by a fixed amount, corresponding to movement on appropriate glide planes, was applied in each iteration. TDs exiting the field at one edge entered at the opposite edge at a random position, and at least one TD had to be glissile for a reaction to take place. The model was exercised for a variety of interaction radii and different mobile:sessile TD ratios, starting with the case with equal populations of all possible Burgers vectors.

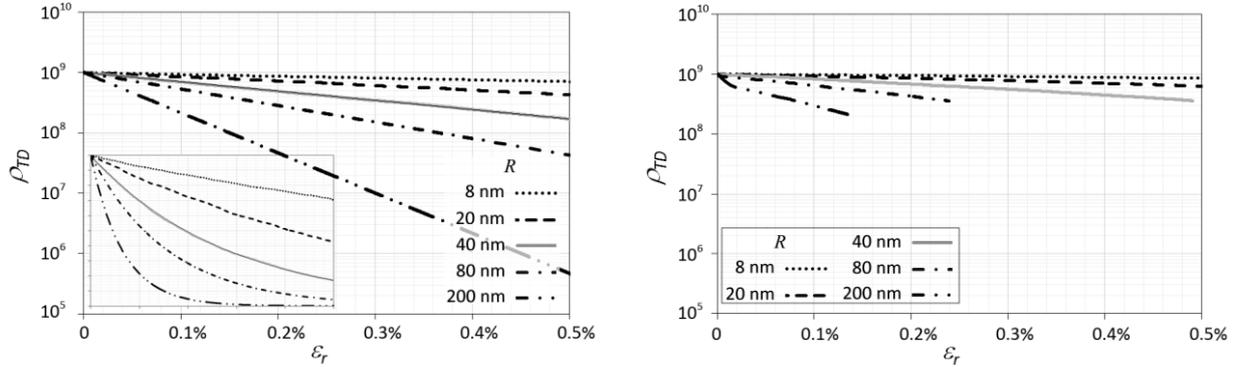

Figure 7. (a) Change in threading dislocation density $\rho_{TD}$ as a function of strain relieved $\varepsilon_r$ for (001) GaAs with an initial $\rho_{TD}$ of $1 \times 10^9$ cm$^{-2}$ and equal populations of all Burgers vectors. The different curves correspond to different interaction radii $R$; the inset shows the same data on a linear scale. (b) An equivalent .plot for (0001) GaN. The termination of curves is due to the elimination of all mobile TDs in the model.

Figure 7 shows the change in total dislocation density as a function of strain relieved $\varepsilon_r$ for fcc cubic and hexagonal systems in this case. For the cubic system, the predictions of the model are encouraging for reduction of threading dislocation density. The key difference between Fig. 7 and Figs. 4-6 is that it is quite feasible to obtain $\varepsilon_r \approx 0.5\%$ in a layer <100 nm – reducing $\rho_{TD}$ by roughly an order of magnitude in a structure many times thinner than the equivalent thick buffer layer. In section 4 we find an interaction radius of $R \sim 40$ nm in GaAs, and one might therefore expect that a stack of five such dislocation filters would reduce $\rho_{TD}$ by 3-4 orders of magnitude in a buffer layer <1μm in thickness. Unfortunately in the case of (0001) GaN the reduction is again limited by the rapid elimination of all mobile TDs, through $2 \rightarrow 2$ reactions that convert mobile TDs to sessile TDs. Furthermore, the reduction that does occur is about half that of GaAs for the same interaction radius, (although estimates of $R$ have been made at around 100nm[39]).

In the models used to produce Fig. 7 the initial populations of dislocations contained equal numbers of all Burgers vectors. For (0001) GaN there seems little point in examining imbalances in the Burgers vector populations in a simple model since the limiting factor – the attrition of mobile TDs – is inescapable. As noted above in the fcc cubic system all $2 \rightarrow 1$ reactions must involve two mobile TDs and one sessile TD. Thus, when two mobile TDs combine they always produce a sessile TD; and when a mobile and a sessile TD combine, they produce a mobile TD. This conversion between mobile and sessile dislocations tends to produce a balance between the two populations and allows the continued reduction of dislocation density. Fig. 8 shows the change in the densities of mobile and sessile TDs for different starting ratios, from 90% to 10% mobile TDs; in every case the densities of the two populations tend to converge as relaxation proceeds.

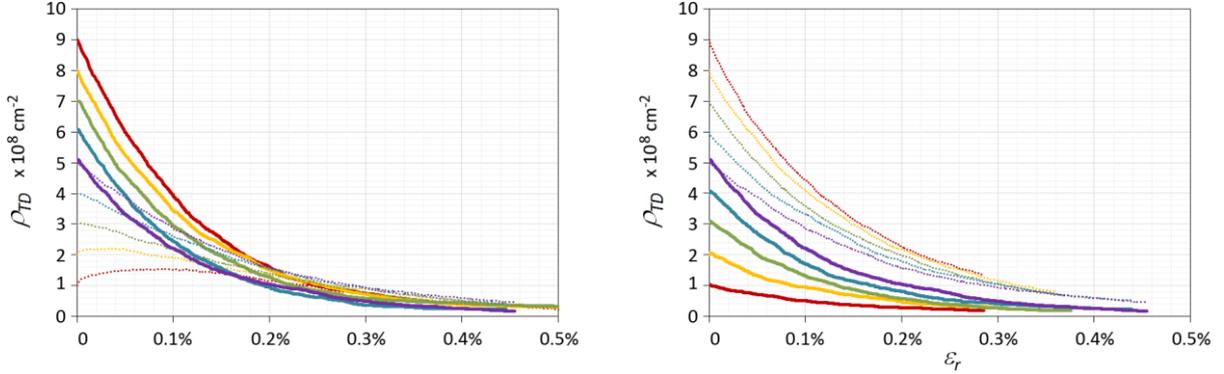

Figure 8. Change in mobile (solid) and sessile (dotted) threading dislocation densities as a function of strain relieved $\varepsilon_r$ for (001) GaAs with an initial $\rho_{TD}^{(0)}$ of $1 \times 10^9$ cm$^{-2}$ and a) initial mobile: sessile ratios from 9:1 to 1:1 and b) 1:1 to 1:9.

In any structure with a decreasing TD density, kinetic constraints will start to become significant as the density of mobile TDs decreases. Although an identical amount of strain relieved $\varepsilon_r$ can be produced with many mobile TDs moving a little, or few TDs moving a lot, if dislocation velocities are the same in both cases it will take longer to produce the same relaxation with fewer mobile TDs – a phenomenon known as dislocation starvation.[47] We may obtain an estimate of the time required to achieve the relaxation levels shown in Fig. 7 – and hence the reduction in $\rho_{TD}$ – by using Eq. (8) and taking a dislocation velocity of ~1μm s$^{-1}$ as an order of magnitude estimate.[48] The result is shown in Fig. 9, and indicates that times of several hours are likely to be required in order to obtain threading dislocation densities significantly below $10^5$ cm$^{-2}$, although the thermally-activated[32] and stress-dependent[19] nature of dislocation glide may give some flexibility to reduce these times significantly in practice. In the case of GaN, there is an upper limit to the reduction in $\rho_{TD}$ (in addition to the kinetic constraints) due to the elimination of mobile TDs. This is consistent with experimental observations of only halving $\rho_{TD}$ even with dislocation filter structures.[41]

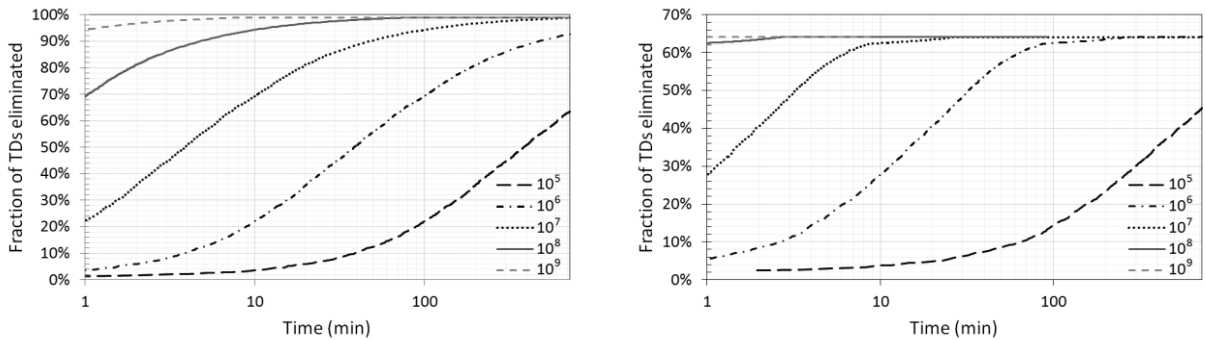

Figure 8. Kinetic limitations to dislocation density reduction caused by finite dislocation velocities

(taken to be ~1μm s$^{-1}$). (a) (001) GaAs and (b) (0001) GaN, $R$ = 40nm with $\rho_{TD}^{(0)}$ from $10^9$ to $10^5$ cm$^{-2}$. Equal proportions of all Burgers vectors were used in both cases.

## 4. MEASUREMENT OF INTERACTION RADIUS

As noted above, the interaction radius $R$ can at best be considered a rough approximation to a complex geometry, where capture of one TD by another occurs over a region that is far from circular and is stress-dependent. Nevertheless, if it is to be used in modelling TD interactions it would be useful to have an indication of its magnitude, since this has a large effect on the efficacy of a dislocation filter layer (Fig. 4). To this end, we have investigated some dislocation filter structures employed in GaAs on Si. The filters consisted of five $In_{0.15}Al_{0.85}As$ layers, 10nm thick, with 10nm GaAs spacer layers and were repeated three times in a buffer layer below a quantum dot laser structure (Fig. 9). The aim of the multilayer structure was to inhibit dislocation multiplication (see discussion below) but for current purposes, the structure can be considered to be equivalent to a single layer with the same total thickness and mean misfit strain, i.e. a layer 90nm thick with composition $In_{0.083}Al_{0.917}As$. This has a strain-thickness product of $\varepsilon h$ = 0.53 nm and in the absence of kinetic effects would be expected to relax to the blocking/trapping limit of $\varepsilon h \approx$ 0.24 nm, i.e. a relaxation of approx. 0.3%.

A typical image showing the effect of the TD filters is shown in Fig. 9. We measured a fractional change, $\rho_{TD} / \rho_{TD}^{(0)}$, of 0.38 ± 0.05 across the dislocation filter structures by taking the sum of all TDs entering and leaving the structures in several micrographs. Comparison with Fig. 7 shows that this corresponds to an effective capture radius $R$ of ~40nm for this system. One might therefore expect a reduction in $\rho_{TD}$ of 1-0.38$^3$ ≈ 95% from a set of three dislocation filter structures.

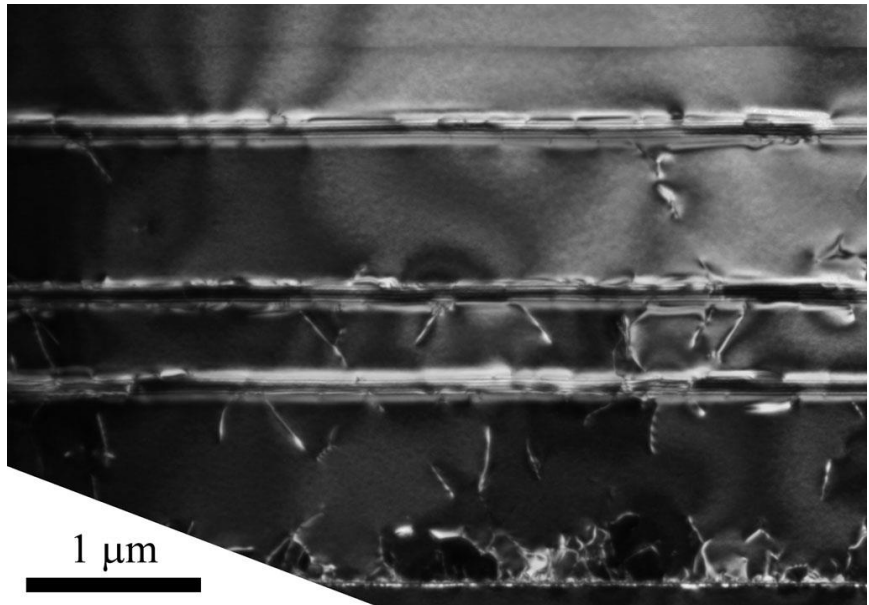

Figure 9. Cross section transmission electron micrograph (dark field, **g** = 004) of a GaAs on Si structure with three dislocation filter structures, producing a reduction in dislocation density of approx. 95%.

## 5. DISCUSSION

Modelling threading dislocations in strained epitaxial layers in an accurate and detailed way remains a formidable task. The details of dislocation interactions *en masse* via their stress fields is starting to be addressed in three-dimensional dislocation dynamics (3DDD)[28,34-36] but there remains much to be done to develop detailed understanding of the processes involved in mismatched epitaxy. The difficulties of crystal growth, with dependencies on temperature, fluxes, chemistry, time and crystallography have not been considered here and these often place tight constraints on what is achievable. However, we have shown that it is possible to make some progress using very simple models that can describe in rough outline the behaviour of dislocations *en masse*. The crystallography must be dealt with properly in order to obtain meaningful results. This is one of the main justifications for the use of a purely numerical, iterative model to generate the main results presented here, since Burgers vectors, interaction radii and dislocation movements are straightforward to incorporate and vary. This is both simpler and gives more flexibility than sets of coupled differential equations,[23,29,30,42,45,49] which in any case require numerical solutions for anything beyond very straightforward cases.

For thick buffer layers, $\rho_{TD}$ has been reported to decrease as $1/h$ in some systems (notably GaAs on Si[26]) rather than $1/h^2$. Dislocation densities are usually found to be of the order of $10^6 – 10^7$ cm$^{-2}$ in layers several μm in thickness,[33] which is actually ten to a hundred times better than might be expected from the models presented in section 2. Nevertheless, we note that the exponential reduction in $\rho_{TD}$ that is a result of dislocation glide appears linear in its initial stages (inset, Fig. 6). It seems likely that the additional misfit stresses that result from differences in thermal expansion coefficient are responsible for increased TD movement and a more rapid decrease in $\rho_{TD}$. This is also consistent with Yamaguchi's[26] study of repeated annealing in GaAs on Si, which showed that $\rho_{TD}$ dropped according to the number of annealing cycles rather than temperature or layer thickness. Zogg[23] proposed that the interaction radius should increase in proportion with thickness for glide in thick layers, rather than being held constant as we used in Eqs. 5-12. This gives a small improvement in $\rho_{TD}$ reduction but does not change the conclusions significantly – thick buffer layers are unlikely to yield very high quality material by themselves and movement of TDs in response to misfit strain is the best way to improve upon this. For example, the use of a reverse-graded Si$_x$Ge$_{1-x}$ layers on top of a thick Ge buffer layer on S was shown to give significant improvements both in $\rho_{TD}$ and surface roughness in comparison with a thick layer alone.[50]

Interestingly, the difference in behaviour between Figs. 4b and 5a shows that the main problem in GaN is not simply the reaction pathway converting two mobile (*a+c*) type dislocations to two sessile or *c*-type *a*-type dislocations. Rather, it is the immobility of sessile TDs that prevents them from interacting with each other, particularly the *a*-type basal plane dislocations. This raises the intriguing possibility that choice of a substrate orientation in which glide forces on pyramidal or rhombohedral planes would allow all dislocation types to move and result in a significant reduction in $\rho_{TD}$.

Our numerical modelling and experimental measurements show that dislocation filter structures in cubic materials should be able to reduce $\rho_{TD}$ by several orders of magnitude, if they can be applied sequentially with similar amounts of relaxation in each dislocation filter. Even better, if all the relaxation (change in lattice constant) can be achieved in the first few such structures, then further filter structures can be optimized for reduction of $\rho_{TD}$ without further change in lattice constant. Moving dislocations around – to-and-fro – at the target lattice constant and at strain-thickness levels below 0.8nm so that no further dislocations are created (nor wanted) is similar to mechanical annealing[16,46,47] and must be the only feasible route to dislocation starvation. In practice, kinetic effects caused by finite dislocation velocities are likely to be the practical limit to what is achievable. Obviously therefore, anything which prevents or inhibits dislocation movement is usually bad. Since dislocation glide is thermally activated,[48] low growth temperatures will inhibit their movement. Planar defects will inhibit, or at least slow down, movement of defects across them. These can include stacking faults or microtwins (e.g. in (111) $Si_{1-x}Ge_x$ on Si[51], where in a dissociated dislocation the force on the leading partial dislocation is higher than that on the trailing partial, driving them apart and producing stacking faults), basal plane stacking faults in non-polar GaN/sapphire[4], or inversion domain boundaries in III-Vs on silicon[52] or GaN[53]. Variations in composition due to phase separation,[54] layer thickness due to surface roughening, or high densities of point defects due to growth far from equilibrium conditions[55] will also act to reduce dislocation movement and may reduce the efficacy of any dislocation filtering structure.

Although it might seem that, in fcc materials, reduction of TD density should be a straightforward matter of maximizing the strain relieved $\varepsilon_r$, this does not take into account dislocation multiplication. If stresses and geometry allow,[56] a dislocation moving in response to the misfit strain may wind around a pinning point to produce a spiral dislocation source – producing two new threading dislocations for every circuit.[12] In cubic layers, this may be necessary for relaxation but is clearly undesirable after the desired lattice cvonstant has been reached and must then be avoided; however in (0001) GaN layers some replenishment of the mobile TD population may actually be beneficial and allow a reduction of the total threading dislocation density. In a layer of constant composition, dislocation multiplication is expected to commence when the strain thickness product exceeds $\sim 4\epsilon_c h_c$. This places an upper limit on the strain and thickness of a dislocation filtering layer, with an increase in $\rho_{TD}$ for layers that are too thick.[26,57] Multiplication can be suppressed by the use of strain across interfaces, restricting the movement of dislocation segments that lie parallel to the surface. Strained layer superlattices or compositionally-modulated structure may thus be expected to give superior performance.

## 6. SUMMARY AND CONCLUSIONS

We have considered the effectiveness of thick buffer layers and threading dislocation filtering structures using simple numerical modelling, correctly taking into account the crystallography of (001) face-centred cubic materials such as $Si_xGe_{1-x}$ and GaAs as well as (0001) GaN. Using TEM of TD filters in GaAs on Si we have measured the effective interaction radius $R$ below which dislocations will react, and obtain a value around 40nm.

For interaction radii of the order of tens of nm we show that simple buffer layers alone are unlikely to produce low threading dislocation densities ($<10^5$ cm$^{-2}$). The key to *significant* reduction in threading dislocation density $\rho_{TD}$ is their movement using misfit stress (i.e. in layers with a strain-thickness product $\varepsilon h > 0.2$ nm, produced either by compositional changes or thermal expansion differences). The most important parameter is the amount of strain relaxed by misfit dislocations in the dislocation filter structure, $\varepsilon_r$. Nevertheless, maximizing $\varepsilon_r$ – while keeping the layer thickness to a minimum – implies the use of as high a misfit strain as possible and there are likely to be constraints imposed by crystal growth that mitigate against this.

In (001) fcc materials there is an exchange between mobile and sessile TDs that leads to a balance between the populations, allowing continued reduction of threading dislocation density $\rho_{TD}$. Although the threading dislocation density reduction is independent of the fraction of TDs that are mobile, finite dislocation mobilities lead to a kinetic limit. This limit may impose relatively long times at high temperatures and require careful design to prevent dislocation multiplication (e.g. $\varepsilon h < 0.8$ nm and use of superlattice structures). In (0001) GaN the conversion of pairs of mobile (*a+c*) type TDs into pairs of sessile TDs is expected to remove almost all mobile TDs, effectively preventing the use of strained layer structures as effective dislocation filters in this material, unless dislocation multiplication can be used to replenish the mobile TD population.

Finally, we reiterate that almost all modelling of this problem to date is rather simplistic and broad brush in nature. We expect that 3DDD modelling,[28,34] in which all interactions are taken into account explicitly, may bring new insights into this complex and technologically important problem.

## 7. ACKNOWLEDGEMENTS

This work was funded by the EPRSC under grants EP/J013048/1 and EP/J012904/1.